\documentclass{aa}       
\usepackage{times,epsf}  

\def\grsim{\,\lower 1mm \hbox{\ueber{\sim}{>}}\,}
\def\lesssim{\,\lower 1mm \hbox{\ueber{\sim}{<}}\,}
\def\ltsima{$\; \buildrel < \over \sim \;$}
\def\simlt{\lower.5ex\hbox{\ltsima}}
\def\gtsima{$\; \buildrel > \over \sim \;$}
\def\simgt{\lower.5ex\hbox{\gtsima}}
\input psfig
\begin{document}

   \thesaurus{03     
              ( 11.03.1; 
                11.09.3; 
                12.03.3; 
                12.03.4; 
                12.04.1; 
                13.25.2)} 

\title{The effect of magnetic fields on the mass determination of
clusters of galaxies} 


   \author{     K. Dolag$^{1}$,              
                S. Schindler$^{2}$          
	}

   \institute{
               MPI f\"ur Astrophysik,                        
               Karl-Schwarzschild-Str. 1,                    
               85748 Garching,                               
               Germany;                                      
                e-mail: {\tt kdolag@mpa-garching.mpg.de}     
\and                                                         
               Astrophysics Research Institute,              
               Liverpool John Moores University,             
	       Twelve Quays House,                           
               Egerton Wharf,                                
               Birkenhead CH41 1LD,                          
               United Kingdom;                               
               e-mail: {\tt sas@staru1.livjm.ac.uk}          
              }                                              
   \date{}

\authorrunning {K. Dolag \& S. Schindler}                    

   \maketitle

   \begin{abstract}
We test the effects of magnetic fields on the mass determination of
galaxy clusters with the X-ray method. 
As this method takes into account only thermal pressure, additional 
non-thermal pressure support, e.g. caused by a magnetic field, can
result in an underestimation of the mass.
SPH models of galaxy clusters 
including magnetic fields of various strengths in two different cosmologies
are used for the test. We compare the true mass
of the model clusters with the mass determined from simulated X-ray images of
the same models. The effect of the magnetic field on the mass 
in relaxed clusters is found to be
negligible compared to other uncertainties, even at small radii, 
where the strongest effects are expected. However in one
model, in which the cluster is in the process of merging, we
find an effect of up to a factor of 2  caused by
the magnetic field.
In relaxed clusters
the mass estimate, averaged over many models, is at 
most 15\% smaller in models with relatively high magnetic field 
of $2\mu$G compared to models without a magnetic field with a small
dependence on the cosmology chosen. 
Therefore we conclude that the presence of a magnetic field
cannot be responsible for the discrepancies of a factor of 2-3 
found between X-ray and lensing mass in the central regions of
relaxed clusters.

   \end{abstract}

%
%

\section{Introduction}

Clusters of galaxies -- the most massive bound objects in the universe
-- are excellent diagnostic tools for cosmological parameters. In
particular, measuring
the masses of clusters provides a way to determine the amount and the
distribution of dark matter and thus $\Omega_\mathrm{m}$. The comparison
with the baryonic component yields an estimate for 
$\Omega_\mathrm{baryon}$, which is another important cosmological parameter,
as current measurements are in contradiction with primordial
nucleosynthesis for a $\Omega=1$ universe (White et al. 1993).

Cluster masses can be measured in various ways, via
the gravitational lensing effect, via the assumption of virial
equilibrium and with X-ray observations using the assumption of
hydrostatic equilibrium. In some clusters these different
methods yield masses which can differ up to a factor of three in the
central region of the cluster with the lensing mass being typically
larger,
e.g. A2218 (Kneib et al. 1995; Squires et al. 1996), MS0440+0204 (Gioia et
al. 1998),  RXJ1347-1145 (Fischer \& Tyson, 1997, 
Schindler et al. 1997; Sahu et al. 1998), Cl0939+4713 (Seitz et
al. 1996; Schindler et al. 1998). Various reasons for these
discrepancies have been suggested.  The presence of substructure and
cluster mergers are obvious ways to explain the differences, but 
 tests with numerical simulations showed that the assumptions of
hydrostatic equilibrium and spherical symmetry are well justified in
relaxed clusters of galaxies (Evrard et al. 1996; 
Schindler 1996) and cannot account for the large discrepancies in
relaxed clusters. 
Also the effects of small-scale density variations were found to be
much smaller than the observed discrepancies (Mathiesen et al. 1999).
As a further possibility for the discrepancies, in particular in the
cluster centres,  Loeb \& Mao (1994)
suggested that the intra-cluster gas might not only be supported by
thermal pressure, but also by considerable magnetic pressure. As
magnetic pressure is not taken into account in the X-ray mass
determination method it potentially 
leads to an underestimation of the mass.


Whereas Loeb \& Mao (1994) suggested that strong magnetic fields can be hidden
to Faraday rotation measurements if the magnetic fields are
tangled, 
Dolag et al. (1999) found in simulations that even clusters with an
overall small magnetic field can be penetrated partially by regions of
high magnetic fields. These simulations are based on a consistent model for
the cluster magnetic field and therefore lead to realistic magnetic
field configurations within the assumed cosmological scenarios. 
Here the magnetic fields strengths and structure is
given by the formation of the cluster in a cosmological
environment. 
Although on average the magnetic pressure in there models 
is much smaller than 
the thermal pressure ($\approx 5\%$ (Dolag et al. 2000b)), 
the models show some scatter around this mean value.
There are domains of high magnetic fields
approaching or sometimes even
exceeding equipartition with the thermal energy. 
The simulations predict the number,
locations and sizes of these domains depending on the
formation history and dynamical state of the cluster,
which provides the possibility to test the importance of these domains. 
So far the influence of these domains on the
mass determination is not clear.

In this work we use the models by Dolag et al. (1999)
to test the influence of such magnetic
fields on the mass determination with the X-ray method.
We compare the true cluster mass,
which is known exactly for the model clusters, with the
mass derived from a simulated X-ray observation of the same model cluster. 

After the description of the method in Sect.~2 we present the results
in Sect.~3 and the conclusions in Sect.~4.

\section{Method}

We use the cosmological MHD code described in Dolag et al.~(1999) to
simulate the formation of magnetised galaxy clusters from an initial
density perturbation field. The code combines the merely gravitational
interaction of a dark-matter component with the hydrodynamics of a
gaseous component. The gravitational interaction of the particles is
evaluated on GRAPE boards (Sugimoto et al. 1990), 
while the gas dynamics is computed in the
SPH approximation (Lucy 1977, Gingold \& Monaghan 1977). It is supplemented with the
magneto-hydrodynamic equations to trace the evolution of magnetic
fields which are frozen into the motion of the gas due to its
assumed ideal electric conductivity. The
back-reaction of the magnetic field onto the gas is included.
The code assumes the intra-cluster medium to be an ideal gas with adiabatic 
index of $\gamma=5/3$. Radiative cooling is not included. As the
surroundings of clusters are important 
because their tidal fields
affect the overall cluster structure and the merging history, 
the simulation volumes 
are surrounded by a layer of
boundary particles whose purpose it is to represent accurately  the
tidal fields in the cluster neighbourhood.
Extensive tests of the code were performed and described in 
 Dolag et al.~(1999). 
Various test problems could be solved successfully, and
$\nabla\cdot{\mathbf B}$ is always negligible compared to the magnetic field
divided by a typical length scale. 

\begin{figure*}
  \centerline{\epsfxsize=0.499\hsize\epsffile{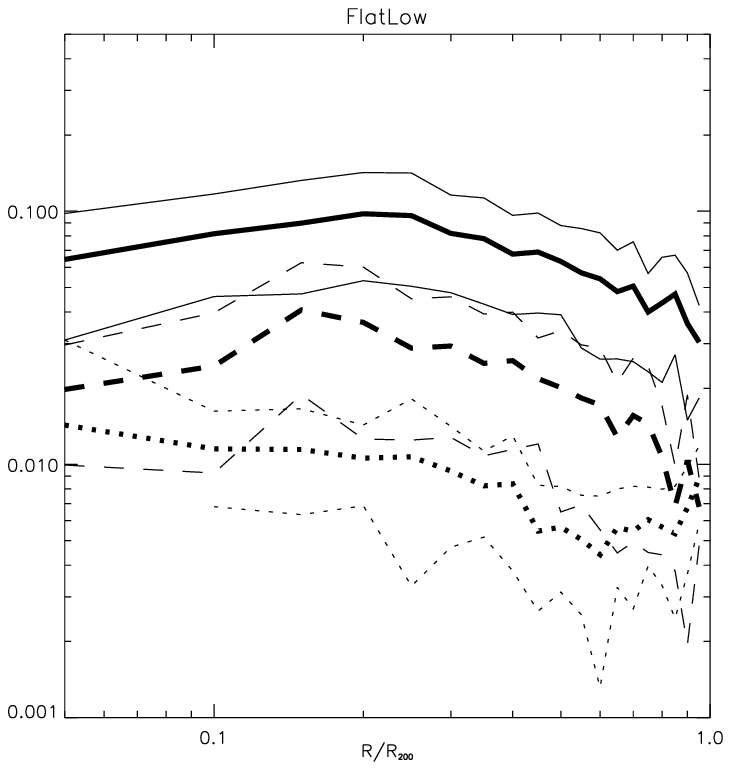} \put(-195.,235.){a} 
              \epsfxsize=0.499\hsize\epsffile{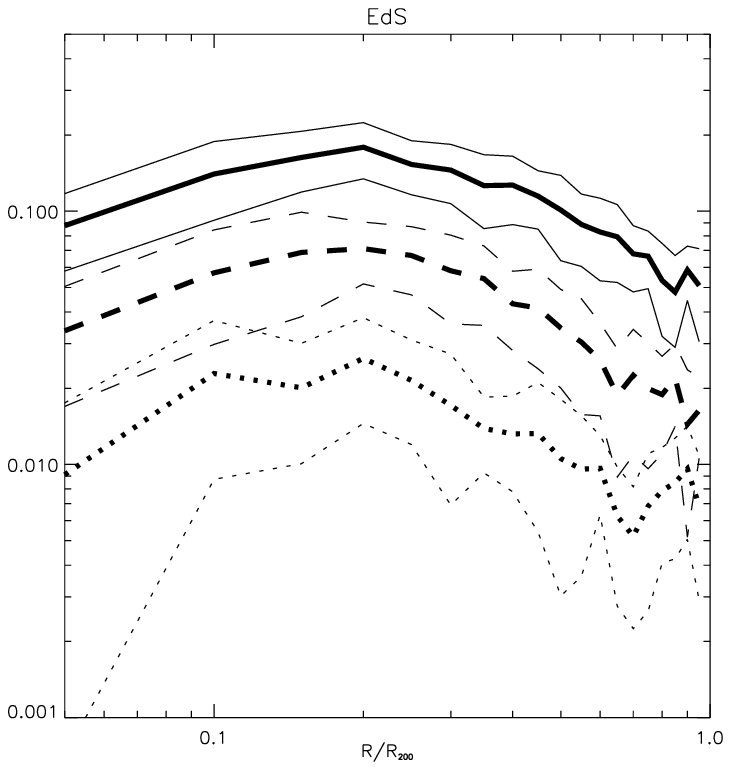}\put(-195.,235.){b}}
\caption[]{Percentage of particles for which the ratio
between magnetic and thermal pressure exceeds of 5\% as function of position
in the cluster (thick lines). Shown is the average 
over all FlatLow models (a) and all
EdS models (b). The different line types distinguish the different
magnetic fields, with the solid line representing a high, the dashed line
a medium and the dotted line a low magnetic field. The thin lines
are the RMSs between the models of the same class.
The radial distance to the cluster centre is in units of the
virial radius R$_{200}$.
}
\label{fig:pers}
\end{figure*}

Two different kinds of cosmological models are used, EdS and FlatLow
(see Table~\ref{tab:models}).
We set up cosmological initial conditions at redshift $z=15$
(for EdS) and $z=20$ (for FlatLow) and follow the formation of the clusters
including the magnetic field. For each
cosmology we calculate ten different realisations which result in clusters 
of different final masses and different dynamical states at redshift 
$z=0$.
 
\begin{table}[t]
\begin{center}
\begin{tabular}{|c|cccccl|}
\hline
 model & $H_0$ & $\Omega_0$ & $\Lambda$ & $\sigma_8$ & $z_0$ & $f_\mathrm{baryon}$ 
\\
\hline
 EdS     & 0.5 & 1.0 & 0.0 & 1.2 & 15 & 5\% \\
 FlatLow & 0.7 & 0.3 & 0.7 & 1.05 & 20 & 10\% \\
\hline
\end{tabular}
\end{center}
\caption{Parameters of the EdS and the FlatLow models.}
\label{tab:models}
\end{table}

\begin{table}[t]
\begin{center}
\begin{tabular}{|c|c|c|c|}
\hline
 model & $B_\mathrm{ini}$ 
       & $\left<B_\mathrm{final}\right>_\mathrm{core}^\mathrm{EdS}$ 
       & $\left<B_\mathrm{final}\right>_\mathrm{core}^\mathrm{FlatLow}$ \\
\hline
no B    & $0.0 \,{\rm G}$          & ---              & --- \\
low B   & $0.2\times10^{-9}\,{\rm G}$ & $0.4\;\mu{\rm G}$ & $0.3\;\mu{\rm G}$\\
medium B& $1.0\times10^{-9}\,{\rm G}$ & $1.1\;\mu{\rm G}$ & $0.8\;\mu{\rm G}$\\
high   B& $5.0\times10^{-9}\,{\rm G}$ & $2.5\;\mu{\rm G}$ & $2.0\;\mu{\rm G}$\\
\hline
\end{tabular}
\end{center}
\caption{Initial magnetic fields (column 2)  and
the resulting final magnetic field strengths in the clusters (column 3
and 4 for EdS and FlatLow models, respectively). 
The final values are an average of the
magnetic field  over the central region (within a radius of 350 kpc) of all
ten clusters for each cosmology. }
\label{tab:bfield}
\end{table}

We simulate each of these clusters with 
different initial
magnetic fields, listed in Table~\ref{tab:bfield}, yielding a total of 80
cluster models.
Since the origin of magnetic fields on cluster scales is unknown, we use
either completely homogeneous or chaotic 
initial
magnetic field structures.
The evolution of the intracluster fields is then computed 
during cluster collapse. 
Previous work 
showed
that (i) the initial field structure is
irrelevant because the final field structure is determined by the
dynamics of the cluster collapse, and (ii) the initial field strengths of
{\it medium B} at 
the starting redshift (see Table~\ref{tab:models})
are adequate to reproduce the Faraday-rotation statistics 
by Kim et al. (1991). The simulated and the observed 
rotation-measure distributions could not be
distinguished significantly by a Kolmogorov-Smirnov test. 
However, the scatter is large because of the 
relatively small number of simulated clusters per set (only 10 cluster
for each cosmology) and because of selection effects
both in the simulations and the
observations.
We therefore allow for a range of initial magnetic field
strengths resulting in a range of synthetic Faraday-rotation
distributions which are all well in agreement with the observed
data. 
The steep power spectrum of these magnetic fields 
suggests that the contribution to the energy density by 
fields tangled on scales smaller than the resolution of the simulation
(\simgt 30kpc)
is negligible.
A detailed description of these models 
and a comparison with rotation measures
is given in Dolag et al. (1999). 
With a realistic model for distributing relativistic electrons
in the simulated clusters, the simulations are also able to match the
properties
of observed radio halos very well (see Dolag \& Ensslin 2000).
As both initial conditions are equivalent for the final clusters we
choose the homogeneous 
initial field set-up, because it is numerically easier to handle.

Due to compression and turbulence in the gas flow, the magnetic 
field is highest in and near the centres of the simulated clusters.
Here the thermal pressure has also the highest values, which leads to 
almost constant ratio of magnetic to thermal pressure support across a
large range of cluster-centric radii, which can amount to 
$\approx 5\%$ for EdS cosmologies and slightly lower values in the
FlatLow cosmologies (Dolag et al. 2000b).
The overall structure of the magnetic field in the outer regions of
the clusters differs from that near the cluster centre. Accretion and
merger events arrange the magnetic fields in the outer cluster regions
in coherent patterns on fairly large scales. Near the cluster centre,
the gas flow pattern is almost randomised, and the magnetic field is
consequently tangled and bent on fairly small scales ($\approx
70$kpc). Although the
magnetic pressure can be considered isotropic at or near the cluster
centre, it may well be anisotropic in the outer parts. Even though the
fraction of the magnetic pressure relative to the thermal pressure, averaged
over spherical shells, is approximately constant across the cluster,
the detailed effects of magnetic pressure on the gas flow depend on
the direction of the magnetic field relative to other locally
preferred directions, like e.g. the orientation of filaments surrounding a
cluster, the local path of matter infall, or the orbit of a merging
sub-clump. Therefore, the same overall magnetic pressure can
have different effects on the local dynamics of gas flows, depending
on whether the magnetic field is ordered on scales comparable to the
cluster scale, and on the orientation of the magnetic field. 
Furthermore in an intra-cluster medium with a small averaged   
fraction of the magnetic relative to the thermal pressure 
may be well regions, in which the magnetic field gives much stronger 
pressure support. Fig.~\ref{fig:pers} shows the mass fraction inside
spherical shells which have more than 5\% magnetic pressure 
support. Note that there is even a small number of regions where the
magnetic pressure is twice as high as the thermal pressure.

In this work we concentrate on the effects of the magnetic
field on the mass determination via the X-ray method. 
Effects like deviation from  hydrostatic equilibrium due 
to mergers,  possible shortcomings of the $\beta$ model and others are not
discussed. In order not to mix up the different
effects we select for the mass tests only the models
not suffering from the above mentioned other effects, i.e. relaxed
clusters. Otherwise
even the average
over many models can be dominated by the strong
deviations of a single model. Furthermore,
a small magnetic field can slightly alter the
dynamical time scale of a simulated cluster. 
Therefore, when comparing simulated clusters with different magnetic
fields at equal redshifts, they can be in slightly different
evolutionary stages according to their own time scale. While this
effect is irrelevant in relaxed clusters it can
change the mass estimate considerably for merging clusters. 

\begin{table}[t]
\begin{center}
\begin{tabular}{|c|c|c|}
\hline
 cosmology & FlatLow & EdS \\ 
\hline
 used clusters & 6/10 & 7/10 \\
\hline
 no B&     $0.98\pm0.31$&$0.85\pm0.32$\\
 low B&    $1.08\pm0.25$&$0.81\pm0.28$\\
 medium B& $0.98\pm0.26$&$0.96\pm0.43$\\
 high B&   $1.02\pm0.22$&$0.77\pm0.32$\\
\hline
\end{tabular}

\end{center}
\caption{Ratio of true mass to X-ray mass at $50 \mathrm{kpc}$ for 4 different
magnetic field strengths in both cosmologies: Mean value and standard
deviation of 21 and 18 model configuration for EdS and FlatLow,
respectively. For each model we use 3 projections.
}
\label{tab:values}
\end{table}

To minimize all 
non-magnetic effects we select 6 out of the FlatLow models
and 7 out of the EdS models, which look round and relaxed
and are not in the process of merging at the
moment of the mass determination ($z=0$).

To calculate X-ray masses from these model
clusters, we project the model clusters onto X-ray images in three
projection directions. We use the ROSAT energy range (0.1-2.4 keV) for
these projections, because for typical cluster temperatures the
observed countrate in this range is almost independent of the temperature. We
calculate maps of projected emission-weighted temperature as well. 
Both projections are binned to $200\times200$ pixels with a resolution of 
20~kpc. The bin size determines the smallest possible radius for our
mass analysis.  We can measure masses only at radii larger than about 2
pixels, i.e. $\approx$ 50~kpc. From X-ray and temperature maps we
calculate surface 
brightness and temperature profiles centred on the X-ray centroid.
We deproject the surface brightness to three-dimensional 
densities with the $\beta$ model
(Cavaliere \& Fusco-Femiano 1976). With the
assumption of hydrostatic equilibrium the integral mass within radius
$r$ can be determined as

$$
M(r) = {-kr\over \mu m_p G} T \left({ d \ln \rho \over d \ln r }+
                                    { d \ln T    \over d \ln r }\right),
   \eqno(1)
$$
with $\rho$ and $T$ being the density and the temperature of the
intra-cluster gas, respectively. 
$k$, $\mu$, $m_p$, and $G$ are the
Boltzmann constant, the molecular weight, the proton mass, and
the gravitational constant.

\section{Results}

\begin{figure*}
  \centerline{\epsfxsize=0.499\hsize\epsffile{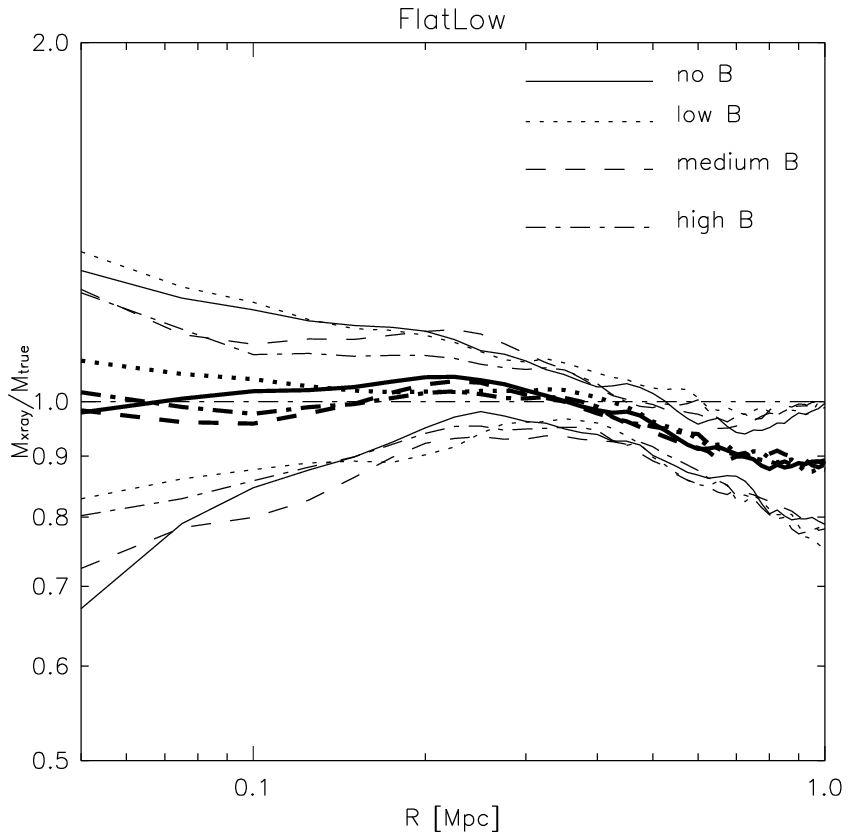} \put(-195.,235.){a} 
              \epsfxsize=0.499\hsize\epsffile{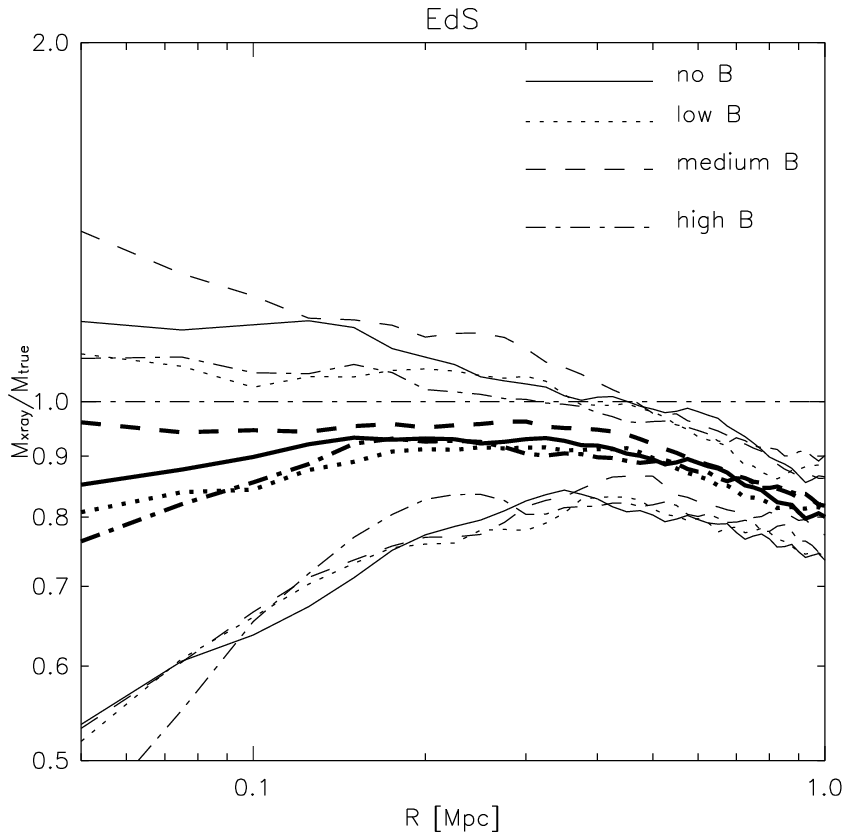}\put(-195.,235.){b}}
\caption[]{Ratio of true and X-ray mass for 4 different
magnetic field strengths in two different cosmologies.
The bold lines are averaged
profiles, while the thin lines show the corresponding standard deviation.  
a) The FlatLow model profiles are 
averaged over 18 configurations (6 models $\times$ 3
projections). b) The EdS model profiles are
averaged over 21 configurations (7 models $\times$ 3
projections). 
In general there are only a marginal trends to underestimate the
mass with increasing magnetic field strength. This effect is much smaller
than the standard deviation and therefore not significant.
}
\label{fig:comp}
\end{figure*}

For the comparison of the true masses of the models 
and the masses derived by the
X-ray method we calculate the ratio of both and plot this ratio versus
radius. 
Figure~\ref{fig:comp}a and b show the comparison for FlatLow and EdS 
models, respectively. The ratio profiles are averaged over various models
and three projection directions (but not over different
initial magnetic fields).

\begin{figure}
  \centerline{\epsfxsize=0.99\hsize\epsffile{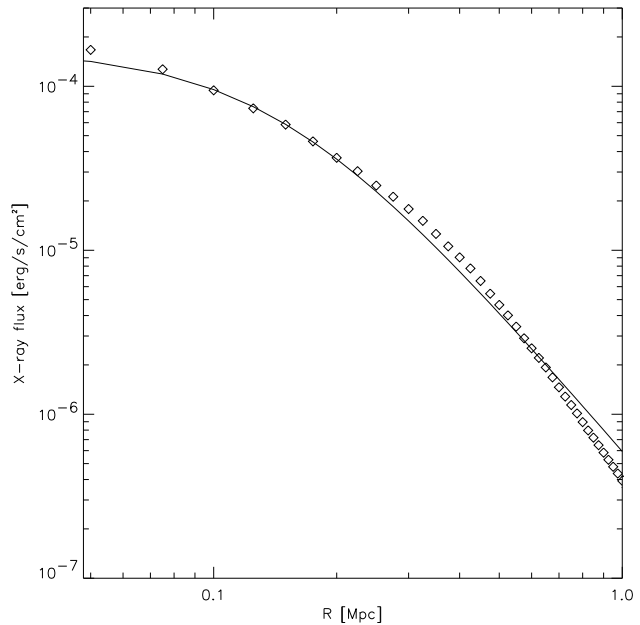}}
\caption[]{X-ray surface brightness profile (diamonds) for one of the relaxed
clusters from the EdS models. The solid line shows the fit of these
data with a $\beta$-model. Obviously, the $\beta$ model is too flat in
the central region as well as at large radii. Therefore the gradients
are underestimated by the $\beta$ model in these regions, which
results in an underestimation of the mass according to Eq. 1.
}
\label{fig:compbeta}
\end{figure}

There is an overall underestimation of the mass visible at very large
and very small
radii, which is not due to the magnetic field, but to
limitations of the $\beta$ model. As shown in Fig.~\ref{fig:compbeta}
the  $\beta$ model cannot fit every part of the profile equally
well. Typically, the $\beta$ model is less steep than the actual profile
in the central region and at large radii. Therefore the density
gradients are underestimated there, which leads to an underestimation of the
mass in these regions according to Eq. (1). A detailed investigation of
different analytic deprojection methods 
will be published in a forthcoming paper.

As expected the effects of the magnetic field are
largest in the central regions 
because the magnetic field is generally stronger in the cluster centre
as seen in simulations (Dolag et. al 1999)  as well as in
observations (Kim et al. 1990; Clark et al. 1999). 
Mostly profiles for high magnetic fields are slightly
below the profiles for no or low magnetic field. The values of the
mass ratios at 50 kpc are listed in
Table~\ref{tab:values}. The difference between the profiles at this
radius is only 10-15\%. For  comparison the standard
deviations of the profiles are shown as well. Clearly, 
the differences 
between the profiles with different magnetic field are much smaller
than the scatter due to other effects like projection effects.
Therefore they are certainly not significant. Moreover, the curves
are not even exactly ordered according to the magnetic field
strength. 
At radii larger than 200 kpc there is no difference at all
between the profiles of models with different magnetic fields.
This comparison of the profiles shows that the mass differences for
different magnetic fields are negligible at all radii.

\begin{figure}
  \centerline{\epsfxsize=0.99\hsize\epsffile{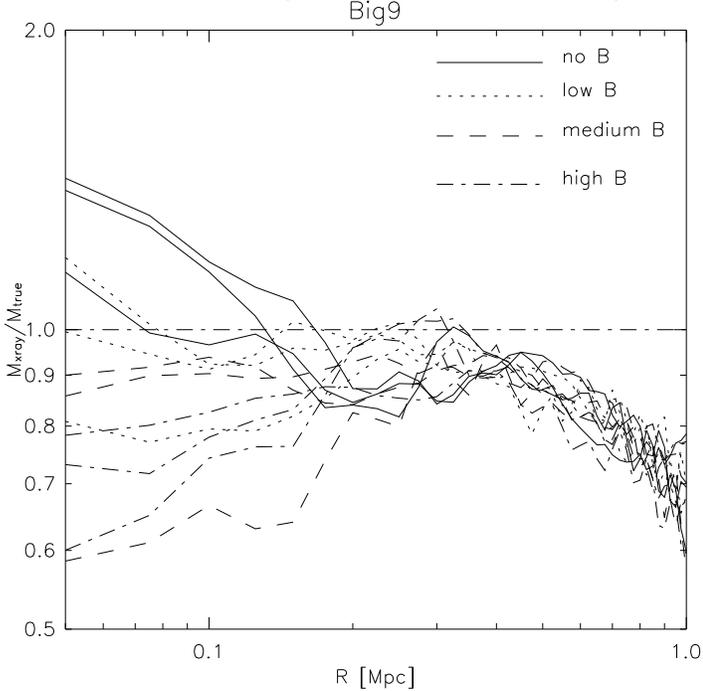}}
\caption[]{Ratio of true and X-ray mass for 
one of the EdS models,
in which the subclusters are in the process of merging. For each
model, the three projection directions are shown in the same
line style. The ordering from over- to underestimating the mass according to
the magnetic field strengths (different line styles) is a result of the combination 
of all effects due to the magnetic field mentioned in the text.
For this extreme merger
the magnetic effects are considerably larger
than the usual scatter of mass profiles from different projection directions.
}
\label{fig:compbig9}
\end{figure}

One of the merger models, which is not used for the profiles in
Fig.~\ref{fig:comp}, shows an interesting effect. The  mass ratios are
clearly ordered 
according to the magnetic field (see  Fig.~\ref{fig:compbig9}).
As it is a merger of two nearly equal subclusters, the mass estimate is very sensitive on
the exact density distribution in the central
region and the positions of the shocks and therefore mass over- or 
underestimations are expected. 
But if these were the only effects one would not expect
the profiles to be ordered according to the magnetic
field as seen in Fig.~\ref{fig:compbig9}. Obviously, the magnetic
field can be important in such extreme merger cases. This is plausible
because magnetic fields
are enhanced during collisions of clusters.

\section{Discussion and Conclusions}

A magnetic field can affect the mass determination in a cluster of galaxies 
in principle in two ways, by the isotropic and the anisotropic component of the
magnetic pressure.
The isotropic component provides an additional non-thermal pressure
support, hence tends to lower
the temperature in the cores of clusters. Depending on the
gas mass fraction, this effect can amount to 
$\approx 5\%$ of the central temperatures for EdS cosmologies (Dolag et
al. 2000). Consequently, the mass is underestimated by the X-ray
method because this additional pressure is not taken into account.
The anisotropic component of the magnetic pressure slightly
changes the density and temperature distribution in the inner region
of the cluster by redirecting gas flows. In this way the gas is
prevented from reaching hydrostatic equilibrium. This may also
lead to a possible misinterpretation of the X-ray 
emission when determining the mass.

To test these effects,
we performed cosmological MHD simulations of galaxy clusters in
different cosmologies, to examine the effect of magnetic fields in
galaxy clusters on the mass reconstruction via the X-ray method.
The range of initial magnetic field strengths was chosen so as to
reproduce the observed Faraday-rotation measurements at the final 
redshift of $z=0$.

We find one extreme merger model, in which the sum of all effects caused
by the magnetic field can reduce 
the reconstructed mass by a factor of 2 in the central region. This
model cluster is in the process of merging and has therefore a
particularly high magnetic field. As the mass determination for
merging clusters canbe unreliable, because they are far
from hydrostatic equilibrium, this adds up to already large
uncertainties, so that we can just emphasise again, that X-ray mass
estimates of merging clusters must be regarded with 
caution. 

In relaxed clusters
magnetic fields  do not generally lead to
any under- or overestimate of 
masses and they
cannot be the dominant reason for the mass 
discrepancy between X-ray mass and lensing mass found in the central region
of some clusters. The 
maximum
difference of averaged mass profiles
between models with high magnetic field and no magnetic field is
15\%, while observational discrepancies between different methods 
of a factor of 2-3 
are
found. It is more likely that
projection effects play an important role. As the gravitational
lensing effect is sensitive to all the mass along the
line-of-sight while the X-ray mass determines only the mass in the
potential well, other mass concentrations along the line-of-sight can
lead to a higher lensing mass compared to the X-ray mass (as
probably seen in Cl0500-24 (Schindler \& Wambsganss 1997)). Another
effect can be that cooling flows lead to wrong determinations of
temperatures when the temperature profile of the cluster is not
available (Allen 1998). Also the $\beta$ model does not always provide
a good fit to the profile at all radii (Bartelmann \& Steinmetz
1996). All these effects will be 
analysed in detail in a forthcoming paper.

\begin{acknowledgements}
We thank
Phil James for carefully reading
the manuscript.
\end{acknowledgements}
%
%

\end{document}